\documentclass[prl,aps]{revtex4}
\usepackage{amsfonts}
\usepackage{amsmath}
\usepackage{amsbsy}
\usepackage{amssymb}

\begin{document}

\title{SOLITON NATURE OF EQUILIBRIUM STATE OF TWO
CHARGED\\[0.3ex] MASSES IN GENERAL RELATIVITY}

\author{G.A.~Alekseev}

\address{\text{Steklov Mathematical Institute RAS, Gubkina 8, Moscow 119991, Moscow, Russia};
International Network of Centers for Relativistic
Astrophysics (ICRANet)\\
\text{Piazzale della Repubblica, 10, 65122 Pescara, Italy}\\
G.A.Alekseev@mi.ras.ru}

\author{V.A.~Belinski}

\address{International Network of Centers for Relativistic
Astrophysics (ICRANet)\\
\text{Piazzale della Repubblica, 10, 65122 Pescara, Italy;}\\
\text{Physics Dept. Rome University "La Sapienza", 00185
Rome, Italy;}\\
Institut des Hautes Etudes Scientifiques (IHES),\\
F-91440 Bures-sur-Yvette, France\\
belinski@icra.it}

\begin{abstract}
New derivation of static equilibrium state for two charged masses in General Relativity is given in the framework of the Inverse Scattering Method as an alternative to our previous derivation of this solution by the Integral Equation Method. This shows that such solution is of solitonic character and represents the particular case of more general (12-parametric) stationary axisymmetric electrovacuum two-soliton solution for two rotating charged objects obtained by one of the authors in 1986. This result gives an additional support to our comprehension that the appropriate analytical continuations of solitonic solutions in the space of their parameters are always possible and that applicability of the Inverse Scattering Method in presence of electromagnetic field is not restricted only to the cases with naked singularities.
\end{abstract}

\maketitle

\section{One-Soliton Solution}
The gravitational solitons as exact solutions of pure gravity Einstein equations have been introduced in the papers \cite{BZ1} and \cite{BZ2} in the framework of the Inverse Scattering Method (ISM). The generalization of this technique for the coupled gravitational and electromagnetic fields was constructed in \cite{A1}. While in pure gravity case this method can be freely applied to produce solutions both of the black hole and the naked singularity types, in the case of the Einstein-Maxwell system the corresponding mathematical procedure can be formulated only in such a way that it generates solutions with naked singularities and not solutions with horizons. Nevertheless, also in this case after one obtains the final form of solution it is possible to continue it analytically in the space of its parameters in order to get the complete family containing solutions with horizons as well. Let's demonstrate this situation in the simplest case of one-soliton solution of the Einstein-Maxwell equations on the Minkowski background which represents the generalized (that is containing also NUT parameter and magnetic charge) Kerr-Newman field.

Metric and electromagnetic potential for the stationary axisymmetric case in cylindrical Weyl coordinates $\left(t,\rho ,z,\varphi \right)$ take the forms:
\begin{gather}
ds^{2}=-f\left(d\rho^{2}+dz^{2}\right) +g_{tt}dt^{2}+2g_{t\varphi
}dt d\varphi +g_{\varphi \varphi }d\varphi ^{2},  \notag \\
\mathcal{A}=\{A_t,0,0,A_\varphi\},\qquad g_{tt}g_{\varphi \varphi }-g_{t\varphi }^{2}=-\rho ^{2}.\label{Weyl1}
\end{gather}
Following the technique of \cite{A1} the generalized Kerr-Newman solution can be obtained by generation of one soliton on the Minkowski background. The corresponding Ernst potentials are:
\begin{equation}
\mathcal{E}=1-\dfrac{2(m-ib)}{R},\qquad \Phi =\dfrac{q+i\mu }{R},\qquad
R=x+ia\,y+m-ib.  \label{KN1}
\end{equation}%
This form of solution is expressed in terms of ellipsoidal ("oblate") coordinates $(x,y)$, related to the Weyl coordinates by the formulas:
\begin{equation}
\rho =\sqrt{x^{2}-\sigma ^{2}}\sqrt{1-y^{2}},\qquad z=z_{0}+xy,
\label{KN2}
\end{equation}
where arbitrary constant $z_{0}$ indicates the location of the source on the axis of symmetry and $\sigma $ is pure imaginary and it is related to other constant parameters of the solution as:
\begin{equation}
\sigma ^{2}=m^{2}+b^{2}-a^{2}-q^{2}\,\mathbf{-}\text{ }\mu ^{2}\leq 0.
\label{KN3}
\end{equation}
For some purposes it is more appropriate to use spheroidal coordinates $\left(r,\theta \right)$ which are connected with Weyl pair $\left( \rho ,z\right)$ by the relations
\begin{equation}
\rho =\left[ \left( r-m\right) ^{2}-\sigma ^{2}\right] ^{1/2}\sin \theta,\qquad z=z_{0}+\left( r-m\right) \cos \theta ,  \label{KN4}
\end{equation}
and with ellipsoidal coordinates $\left( x,y\right) $ by formulas:
\begin{equation}
x=r-m,\qquad y=\cos \theta .  \label{KN5}
\end{equation}
The conformal factor $f$ in (\ref{Weyl1}) has the following form:
\begin{equation}
f=c_{0}\dfrac{R\bar{R}}{(x^{2}-\sigma ^{2}y^{2})},\qquad R=x+ia\,y+m-ib
\label{KN6}
\end{equation}
where $c_{0}$ is arbitrary constant. Here and in the sequel the bar over a letter means complex conjugation.

The point is that the generating algorithm of \cite{A1} admits the parameter $\sigma $ to be pure imaginary only, i.e. it works provided $\sigma ^{2}\leq 0$. This means that solution (\ref{KN1}) coincides with that part of the generalized Kerr-Newman family which describes not a black hole, but a naked singularity with the mass $m$, NUT-parameter $b$, angular momentum $a$ , electric charge $q$ and magnetic charge $\mu $. Thus, the solution contains five free physical parameters. (The constant $z_{0}$ is not consider as essential parameter due to the trivial freedom of translation along the symmetry axis, and the arbitrary constant $c_{0}$ in the metric coefficient $f$ also does not represents a
physical degree of freedom because usually we fix it to provide the standard asymptotic flatness condition, that is, in particular, $f=1$ at spatial infinity ${}^{\text{\footnotemark}}$\footnotetext{In some more complicate cases, such as e.g., two interacting sources, the parameter $c_0$ can play physically more important role: its appropriate choice can change the solution with struts separating the sources to the solution with strings which stretch out along the external parts of the axis of symmetry from the sources to spatial infinity.}).

In the case under consideration the formal constraint of the ISM for parameter $\sigma $ to be pure imaginary generates no real restrictions. It is easy to see, that this parameter can enter the expressions for the metric and electromagnetic potential only in even powers and in such a way that these expressions remain real and correspond to the physical signature of the space-time even if $\sigma ^{2}>0.$ Because of that, in the final form of solution (\ref{KN1}) the restriction $m^{2}+b^{2}-a^{2}-q^{2}\,-$ $\mu
^{2}\leq 0$ on the range of the arbitrary constants can be relaxed. This trivial extension of the range of parameters (corresponding to the transformation $\sigma \rightarrow i\sigma $) can be considered as analytical continuation of this solution in the space of parameters. Thus we obtain from our one-soliton solution for the field of a naked singularity the complete family of generalized Kerr-Newman solutions which includes also the black hole part.

\section{General Two-Soliton Solution For Two Rotating Charged Masses}

The stationary axisymmetric two-soliton solution on the Minkowski background which describes the interaction of the fields of two sources of the generalized Kerr-Newman type have been obtained in explicit form in \cite{A2} (its more detailed exposition can be found also in \cite{A3}). Again, originally this solution emerged from the ISM as corresponding formally to naked singularities but it also admits the similar analytical continuation which allows to obtain the parts of this solution which describe the interaction of fields of a naked singularity and a black hole as well as of two black holes. However, more complicate character of two-soliton solution makes this analytical continuation not so obvious and trivial. The Ernst potentials of this solution are:
\begin{equation}
\mathcal{E}=1-\dfrac{2(m_{1}-ib_{1})}{R_{1}}-\dfrac{2(m_{2}-ib_{2})}{R_{2}}%
,\qquad \Phi =\dfrac{q_{1}+i\mu _{1}}{R_{1}}+\dfrac{q_{2}+i\mu _{2}}{R_{2}},
\label{DKN1}
\end{equation}
\begin{eqnarray}
\dfrac{1}{R_{1}} &=&\dfrac{x_{2}+ia_{2}y_{2}+K_{1}(x_{2}-\sigma
_{2}y_{2})+L_{1}(x_{1}+\sigma _{1}y_{1})}{D},  \label{DKN2} \notag\\
\dfrac{1}{R_{2}} &=&\dfrac{x_{1}+ia_{1}y_{1}+K_{2}(x_{1}-\sigma
_{1}y_{1})+L_{2}(x_{2}+\sigma _{2}y_{2})}{D},
\end{eqnarray}
\begin{eqnarray}
D &=&(x_{1}+ia_{1}y_{1}+m_{1}-ib_{1})(x_{2}+ia_{2}y_{2}+m_{2}-ib_{2})-
\label{DKN3}   \notag\\
&&-\left[ m_{1}-ib_{1}-K_{2}(x_{1}-\sigma _{1}y_{1})-L_{2}(x_{2}+\sigma
_{2}y_{2})\right] \times  \notag \\
&&\times \left[ m_{2}-ib_{2}-K_{1}(x_{2}-\sigma
_{2}y_{2})-L_{1}(x_{1}+\sigma _{1}y_{1}\right] ,
\end{eqnarray}
\begin{eqnarray}
K_{1} &=&\dfrac{ia_{2}-\sigma _{2}}{\sigma _{1}+\sigma _{2}+l},\quad L_{1}=\dfrac{
(m_{1}+ib_{1})(m_{2}-ib_{2})-\left( q_{1}-i\mu _{1}\right) \left( q_{2}+i\mu
_{2}\right) }{(ia_{1}-\sigma _{1})(\sigma _{1}+\sigma _{2}+l)},  \label{DKN4}\notag
\\
K_{2} &=&\dfrac{ia_{1}-\sigma _{1}}{\sigma _{1}+\sigma _{2}-l},\quad L_{2}=\dfrac{%
(m_{1}-ib_{1})(m_{2}+ib_{2})-\left( q_{1}+i\mu _{1}\right) \left( q_{2}-i\mu
_{2}\right) }{(ia_{2}-\sigma _{2})(\sigma _{1}+\sigma _{2}-l)},
\end{eqnarray}
\begin{eqnarray}
\sigma _{1}^{2} &=&m_{1}^{2}+b_{1}^{2}-a_{1}^{2}-q_{1}^{2}-\mu _{1}^{2}\leq
0,  \label{DKN5}\notag \\
\sigma _{2}^{2} &=&m_{2}^{2}+b_{2}^{2}-a_{2}^{2}-q_{2}^{2}-\mu _{2}^{2}\leq
0,
\end{eqnarray}%
\begin{equation}
l=z_{2}-z_{1}.  \label{DKN6}
\end{equation}%
In the last formula, the arbitrary constants $z_{1}$ and $z_{2}$ correspond to locations of the sources on the symmetry axis and $l$ is a measure of the distance between them. The conformal factor $f$ for this solution has the following form:
\begin{equation}
f=c_{0}\dfrac{D\bar{D}}{(x_{1}^{2}-\sigma
_{1}^{2}y_{1}^{2})(x_{2}^{2}-\sigma _{2}^{2}y_{2}^{2})}.  \label{DKN7}
\end{equation}
Now we have two pairs of spheroidal coordinates $\left( r_{1},\theta_{1}\right) $ and $\left( r_{2},\theta _{2}\right) $ which are centered at each source. These pairs of variables are connected with original Weyl coordinates $\left( \rho ,z\right) $ of the metric (\ref{Weyl1}) by the
formulas:
\begin{eqnarray}
\rho &=&\left[ \left( r_{1}-m_{1}\right) ^{2}-\sigma _{1}^{2}\right]
^{1/2}\sin \theta _{1},\qquad z=z_{1}+\left( r_{1}-m_{1}\right)
\cos \theta _{1},  \label{DKN8}\notag \\
\rho &=&\left[ \left( r_{2}-m_{2}\right) ^{2}-\sigma _{2}^{2}\right]
^{1/2}\sin \theta _{2},\qquad z=z_{2}+\left( r_{2}-m_{2}\right)
\cos \theta _{2}.
\end{eqnarray}
Accordingly, for two pairs of ellipsoidal coordinates $\left( x_{1},y_{1}\right)
$ and $\left( x_{2},y_{2}\right) $, in terms of which the
solution (\ref{DKN1})-(\ref{DKN7}) is expressed, we have:
\begin{eqnarray}
x_{1} &=&r_{1}-m_{1},\qquad y_{1}=\cos \theta _{1},  \label{DKN9}\notag \\
x_{2} &=&r_{2}-m_{2},\qquad y_{2}=\cos \theta _{2}.
\end{eqnarray}

Taking into account the distant $l$ we evaluate the total number of free physical parameters in the solution to be 12. The constants $m_{i}$, $b_{i}$, $a_{i}$, $q_{i}$, $\mu _{i}$ ($i=1,2)$ correspond to the presence of masses, NUT-parameters, angular momentums, electric and magnetic charges of two sources.
However, this should not give rise to some confusion
about the physical interpretation of these parameters. On one hand, it is easy to see that if we shift one of the sources (say, the second one, whose parameters are denoted by the suffix "$2$") to a large distance keeping all other parameters finite, we find that in this limit our two-soliton solution coincides with one-soliton solution, that is with the generalized Kerr-Newman solution with the mass $m_{1}$, NUT-parameter $b_{1}$, angular
momentum $a_{1}$ and the electric and magnetic charges $q_{1,}$ $\mu _{1}$. To see this, it is necessary to observe that in this limit we should consider $l\rightarrow \infty $, $(x_{2}/\ell)\rightarrow 1$ and $y_{2}\rightarrow -1$, while the coordinates $x_{1}$ and $y_{1}$ remain finite. However, it is important to emphasize, that in the case of interacting sources (when the distance between them is finite) the parameters mentioned above become pure formal quantities which do not play more the roles respectively of masses, NUT-parameters, angular momentums and charges of each of the sources, and the values of the
corresponding physical parameters should be calculated as functions (in general rather complicated) of the mentioned above formal constants.

\section{A Subtlety}

The 12-parametric two-soliton solution presented in previous section follows from the ISM algorithm again under the condition that both $\sigma _{1}$ and $ \sigma _{2}$ should be pure imaginary. However, now the technical snag is that these two quantities enter the expressions (\ref{DKN1})-(\ref{DKN7})
not only in even powers and in much more intricate way with respect to what we had for one-soliton case. Then the simple replacement $\sigma \rightarrow i\sigma $ as a way to obtain solutions containing also black holes does not work. Besides this  difficulty, it was not clear how one can extract from the expressions (\ref{DKN1})-(\ref{DKN7}) the particular case
of static two-soliton configuration.

This was the reason why we derived in \cite{A4} the static configuration of two interacting Reissner-Nordstrom sources in mutual equilibrium not with the help of result (\ref{DKN1})-(\ref{DKN7}) but using another form of solution
for interacting charged masses which had been found in \cite{A5} (see also Appendix in \cite{A6}) for the general case of N axially symmetric stationary sources using the Monodromy Transform Approach (the Integral Equation Method) constructed in \cite{A7}. How one has to act to derive from these results the most simple form of the solution for the particular two-body case N=2, we explained in details in \cite{A8}. In general, this static asymptotically flat two-body system contains five arbitrary parameters${}^{\text{\footnotemark}}$\footnotetext{We chose the parameter $c_0$ so that the solution becomes asymptotically flat and in particular, $f\rightarrow 1$ at spatial infinity.}, and the first remarkable and surprising for us point  was that this solution simplified drastically when we chose for its five independent parameters the physical masses and charges of the sources together with their separating distance (see \cite{A8} for details).

However, in this 5-parametric solution, the "equilibrium" of the configuration emerges due to the trivial reason, namely due to the conic singularities on the symmetry axis between the sources which constitute the strut preventing the charged massive bodies to fall on or run away from each other. This was not yet a result we had been interested in. The actual equilibrium configuration should be free of any extraneous struts and it was the next remarkable point that we indeed found such configurations after eliminating the conic singularity from the aforementioned 5-parametric solution by imposing only one constraint on its physical parameters. After we  found how to send the singularity away, the next simplification of the solution happened, and just this final extremely simple form of the solution containing 4 free physical parameters we presented in \cite{A4}. It turned out that the equilibrium is possible only between black hole and naked singularity, while the configurations black hole - black hole and naked singularity - naked singularity can not stay in equilibrium without extraneous support.

Intuitively it is clear that both the above mentioned 5-parametric and 4-parametric solutions must be of soliton nature in the sense of the ISM and both must be particular cases of the most general 12-parametric soliton solution (\ref{DKN1})-(\ref{DKN9}). However, up to now we had no a simple and direct proof. The purpose of this paper is to provide this proof by showing the direct mathematical procedure how one can derive these static 5-parametric and equilibrium 4-parametric solutions from the old result (\ref{DKN1})-(\ref{DKN9}). By the way, this proof also gives an evidance in favour of a general statement (as we already
mentioned above) that electrovacuum solutions constructed in the framework of the ISM, proposed in \cite{A1}, similar to vacuum solitons \cite{BZ1,BZ2}, remain valid also for the configurations with horizons and it is pure technical task to find an efficient way of their analytical continuation in the space of parameters.

\section{Static Limit of the General Two-Soliton Solution}

It turns out (and this is the main finding of the present paper) that the 12-parametric two-soliton solution (\ref{DKN1})-(\ref{DKN9}) can be reduced
to the general asymptotically flat 5-parametric static subfamily using the following substitutions:
\begin{eqnarray}
a_{1} &=&\dfrac{\gamma ^{2}}{s_{2}},\quad b_{1}=\dfrac{q_{2}}{s_{2}}\gamma,\quad\mu _{1}=-\dfrac{m_{2}}{s_{2}}\gamma,\quad\sigma _{1}=is_{1},
\label{stat1} \notag\\
a_{2} &=&\dfrac{\gamma ^{2}}{s_{1}},\quad b_{2}=\dfrac{q_{1}}{s_{1}}\gamma,\quad\mu _{2}=-\dfrac{m_{1}}{s_{1}}\gamma,\quad\sigma _{2}=is_{2},
\end{eqnarray}
where
\begin{equation}
\gamma =\frac{m_{2}q_{1}-m_{1}q_{2}}{l}.  \label{stat2}
\end{equation}
The other five real parameters $l,$ $m_{1},\,m_{2},\,q_{1},\,q_{2}$ are the same and they become free independent parameters of the solution. To perform the reduction (\ref{stat1}), we make at first the following transformations. Let us represent the Ernst potentials (\ref{DKN1}) in the form:
\begin{equation}
\mathcal{E}=\dfrac{\mathcal{D}-\mathcal{G}}{\mathcal{D}+\mathcal{G}},\qquad
\Phi =\dfrac{\mathcal{F}}{\mathcal{D}+\mathcal{G}}.  \label{stat3}
\end{equation}
For this we introduced here the functions $\mathcal{D}$, $\mathcal{G}$ and $ \mathcal{F}$ which are homogeneous polynomials of the coordinates $(x_{1},x_{2},y_{1},y_{2})$. In particular, the sum $\mathcal{D}+\mathcal{G}$ coincides up to a constant coefficient with the polynomial $D$ in (\ref{DKN3}). The corresponding constant coefficient we choose for simplicity in such a way that the coefficient in the polynomial $\mathcal{D}$ in front of the monomial $x_{1}x_{2}$ would be equal to $1$. It is easy to see from the expressions (\ref{DKN3}) that this constant multiplier is equal to $(1-K_{1}K_{2}-L_{1}L_{2})^{-1}$. Thus the homogeneous polynomials $\mathcal{D}$ and $\mathcal{G}$ for the 12-parametric two-soliton solution are defined by the relation
\begin{equation}
\mathcal{D}+\mathcal{G}=\dfrac{D}{1-K_{1}K_{2}-L_{1}L_{2}}  \label{stat4}
\end{equation}
and by a supplement condition that $\mathcal{D}$ is quadratic and $\mathcal{G}$ is linear polynomials of the ellipsoidal coordinates $(x_{1},x_{2},y_{1},y_{2})$. Then, comparing the expressions for $\Phi $ in (\ref{stat3}) and (\ref{DKN1}), it is not difficult to read off the expression for $\mathcal{F}$ which is a linear polynomial of ellipsoidal coordinates.

Now we are ready to make the substitution (\ref{stat1}) into the polynomials $\mathcal{D},$ $\mathcal{G}$ and $\mathcal{F}$. The result can be written in a rather compact form as
\begin{eqnarray}
\mathcal{D} &=&x_{1}x_{2}-\gamma ^{2}y_{1}y_{2}+  \label{stat6}\notag \\
&&+\delta \left[ x_{1}^{2}+x_{2}^{2}-\sigma _{1}^{2}y_{1}^{2}-\sigma
_{2}^{2}y_{2}^{2}+2(m_{1}m_{2}-q_{1}q_{2})y_{1}y_{2}\right] ,  \notag \\
\mathcal{G} &=&m_{1}x_{2}+m_{2}x_{1}+\gamma (q_{1}y_{1}+q_{2}y_{2})+  \notag
\\
&&+2\delta \left[ m_{1}x_{1}+m_{2}x_{2}+y_{1}(q_{2}\gamma
-m_{1}l)+y_{2}(q_{1}\gamma +m_{2}l)\right] ,  \notag \\
\mathcal{F} &=&q_{1}x_{2}+q_{2}x_{1}+\gamma (m_{1}y_{1}+m_{2}y_{2})+  \notag
\\
&&+2\delta \left[ q_{1}x_{1}+q_{2}x_{2}+y_{1}(m_{2}\gamma
-q_{1}l)+y_{2}(m_{1}\gamma +q_{2}l\right] ,
\end{eqnarray}
where parameters $\delta ,$ $\sigma _{1}^{2},$ $\sigma _{2}^{2}$ are:
\begin{equation}
\delta =\dfrac{m_{1}m_{2}-q_{1}q_{2}}{%
l^{2}-m_{1}^{2}-m_{2}^{2}+q_{1}^{2}+q_{2}^{2}}.  \label{stat7}
\end{equation}%
\begin{equation}
\sigma _{1}^{2}=m_{1}^{2}-q_{1}^{2}+\gamma ^{2},\qquad \sigma
_{2}^{2}=m_{2}^{2}-q_{2}^{2}+\gamma ^{2}.  \label{stat5}
\end{equation}%
For coordinates we have the same expressions (\ref{DKN8})-(\ref{DKN9}) but
now with $\sigma _{1}^{2}$ and $\sigma _{2}^{2}$ from (\ref{stat5}). The polynomials (\ref{stat6}) determine the static solution with the Ernst potentials (\ref{stat3}).

In the standard way we can find the metric and electromagnetic field corresponding to the Ernst potentials (\ref{stat3})-(\ref{stat5}). These are:
\begin{equation}
ds^{2}=-f(d\rho ^{2}+dz^{2})+H dt^{2}-\dfrac{\rho ^{2}}{H}d\varphi ^{2},
\label{stat8}
\end{equation}
\begin{equation}
A_{t}=\Phi ,\qquad A_{\rho }=A_{z}=A_{\varphi }=0,  \label{stat9}
\end{equation}
where
\begin{equation}
H=\dfrac{\mathcal{D}^{2}-\mathcal{G}^{2}+\mathcal{F}^{2}}{(\mathcal{D}+%
\mathcal{G})^{2}},\qquad \Phi =\dfrac{\mathcal{F}}{\mathcal{D}+\mathcal{G}}%
,\qquad f=\dfrac{c_{0}(\mathcal{D}+\mathcal{G})^{2}}{(x_{1}^{2}-\sigma
_{1}^{2}y_{1}^{2})(x_{2}^{2}-\sigma _{2}^{2}y_{2}^{2})},  \label{stat10}
\end{equation}%
and polynomial functions $\mathcal{D}$, $\mathcal{G}$, $\mathcal{F}$ are
given by the expressions (\ref{stat6}).

We choose the arbitrary constant $c_{0}$ in the metric coefficient $f$ to provide the asymptotic flatness condition, that is $f\rightarrow 1$ at spatial infinity. It is easy to show that for the static solution (\ref{stat6})-(\ref{stat10}) this condition lead to the relation:
\begin{equation}
c_{0}=\dfrac{1}{(1+2\delta )^{2}}.  \label{stat11}
\end{equation}
It is easy to see that the expressions (\ref{stat3})-(\ref{stat11}) are identical to those derived in \cite{A8} from the Monodromy Transform Approach. It is necessary to emphasize that here we derived this solution as a static subfamily of general 12-parametric two\textit{-soliton} solution, and, as far as we started with ISM, this static solution formally is valid only for the case of two naked singularities. However, similarly to the one-soliton case considered above, the static solution derived in this section does not depend on any odd powers of parameters $\sigma _{1}$ and $\sigma _{2}$ which real or imaginary values
determine the character of the source (a black hole or a naked singularity). This allows to relax the restrictions on these parameters to be pure imaginary and to extend trivially our static two-soliton solution to a complete family of static solutions which includes all possible combinations of the Reissner-Nordström sources, that is a pair of black holes, a pair
of naked singularities or a mixed pair.

\section{Static True Equilibrium State For Two Charged Objects}

The real equilibrium configurations of two charged sources (found originally in \cite{A4}) arise from the static solution described in the previous sections after the additional requirement of absence of struts (conical singularities) on the axis of symmetry which implies the condition $\delta =0$. In this case we have
\begin{equation}
m_{1}m_{2}=q_{1}q_{2}  \label{PR1}
\end{equation}%
and under this constraint the polynomials $\mathcal{D}$, $\mathcal{G}$, $%
\mathcal{F}$ take much more simple form:%
\begin{eqnarray}
\mathcal{D} &=&x_{1}x_{2}-\gamma ^{2}y_{1}y_{2},  \label{PR2}\notag \\
\mathcal{G} &=&m_{1}x_{2}+m_{2}x_{1}+\gamma (q_{1}y_{1}+q_{2}y_{2}),  \notag
\\
\mathcal{F} &=&q_{1}x_{2}+q_{2}x_{1}+\gamma (m_{1}y_{1}+m_{2}y_{2}).
\end{eqnarray}%
For these equilibrium configurations the metric coefficients and electric potential in (\ref{stat8})-(\ref{stat9}) also can be expressed in more explicit form by substituting the polynomials (\ref{PR2}) into the expressions (\ref{stat10}).
However, this static equilibrium solution in our first paper \cite{A4} was presented in spheroidal coordinates (\ref{DKN8})-(\ref{DKN9}) as:
\begin{eqnarray}
H &=&\dfrac{[(r_{1}-m_{1})^{2}-\sigma _{1}^{2}+\gamma ^{2}\sin ^{2}\theta_{2}][(r_{2}-m_{2})^{2}-\sigma _{2}^{2}+\gamma ^{2}\sin ^{2}\theta _{1}]}{[r_{1}r_{2}-(e_{1}-\gamma -\gamma \cos \theta _{2})(e_{2}+\gamma -\gamma
\cos \theta _{1})]^{2}},  \label{PR3}\notag \\
\Phi &=&\dfrac{(e_{1}-\gamma )(r_{2}-m_{2})+(e_{2}+\gamma
)(r_{1}-m_{1})+\gamma (m_{1}\cos \theta _{1}+m_{2}\cos \theta _{2})}{r_{1}r_{2}-(e_{1}-\gamma -\gamma \cos \theta _{2})(e_{2}+\gamma -\gamma \cos
\theta _{1})},  \notag \\
f &=&\dfrac{[r_{1}r_{2}-(e_{1}-\gamma -\gamma \cos \theta_{2})(e_{2}+\gamma-\gamma \cos \theta _{1})]^{2}}{[(r_{1}-m_{1})^{2}-\sigma _{1}^{2}\cos
^{2}\theta _{1}][(r_{2}-m_{2})^{2}-\sigma _{2}^{2}\cos ^{2}\theta _{2}]}.
\end{eqnarray}
In this form instead of the formal charge parameters $q_{1},q_{2}$ we used the real physical electric charges of the sources $e_{1}$ and $e_{2}:$
\[
e_{1}=q_{1}+\gamma ,\text{ }e_{2}=q_{2}-\gamma .  \label{PR4}
\]
In terms of physical charges the quantities $\gamma ,$ $\sigma_{1}$, $\sigma _{2}$ entering expressions (\ref{PR3}) are:
\[
\gamma =(m_{2}e_{1}-m_{1}e_{2})(l+m_{1}+m_{2})^{-1},  \label{PR5}
\]
\[
\sigma _{1}^{2}=m_{1}^{2}-e_{1}^{2}+2e_{1}\gamma ,\quad \sigma_{2}^{2}=m_{2}^{2}-e_{2}^{2}-2e_{2}\gamma  \label{PR6}
\]
and additional condition (\ref{PR1}) eliminating the struts take now the following form:
\[
m_{1}m_{2}=\left( e_{1}-\gamma \right) \text{ }\left( e_{2}+\gamma \right) .
\]
While the constants $e_{1}$ and $e_{2}$ give directly the physical charges of each individual source, the parameters $m_{1}$ and $m_{2}$ represent the total relativistic energy of each source in the external field produced by its partner, see details in \cite{A4}.

\section*{Acknowledgments}
The work of GAA was supported in parts by the Russian Foundation for Basic Research (grants 11-01-00034, 11-01-00440) and the program "Mathematical Methods of Nonlinear Dynamics" of the Russian Academy of Sciences.

\end{document}